\documentclass{nature}

\usepackage{graphicx}
\makeatletter
\let\saved@includegraphics\includegraphics
\AtBeginDocument{\let\includegraphics\saved@includegraphics}
\renewenvironment*{figure}{\@float{figure}}{\end@float}
\makeatother

\usepackage{amssymb}
\usepackage{yfonts}

\title{Scattering Assisted Imaging}

\author{ Marco Leonetti$^{1,2}$,Alfonso Grimaldi$^{1}$, Silvia Ghirga$^{1,3}$,  Giancarlo Ruocco$^{1,3}$, Giuseppe Antonacci$^{1}$   }

\begin{document}

\maketitle

\begin{affiliations}
 \item  Center for Life Nano science @ Sapienza, Isituto Italiano di Tecnologia, Viale Regina Elena, 291, I-00161 Roma, Italia.
 \item  CNR NANOTEC-Institute of Nanotechnology c/o Campus Ecotekne, University of Salento, Via Monteroni, 73100 Lecce.
 \item  Dipartimento di Fisica, Universit\'{a} ``La Sapienza'', Piazzale Aldo Moro, 5, I-00185 Roma, Italia

\end{affiliations}

\begin{abstract}
An ideal imaging system provides a spatial resolution that is ultimately dictated by the numerical aperture (NA) of the illumination and collection optics\cite{abbe1873beitrage,rayleigh1903theory}. In biological tissue, resolution is further affected by scattering\cite{sheng2006introduction,bertolotti2012non,vellekoop2010exploiting,park2013subwavelength}  limiting the penetration depth to a few tenths of microns. Here, we exploit the properties of speckle patterns embedded into a strongly scattering matrix to generate a high-resolution illumination. Combining adaptive optics with a custom deconvolution algorithm, we obtain an increase in the transverse spatial resolution by a factor of 2.5 with respect to the diffraction limit. This Scattering Assisted Imaging (SAI) is compatible with long working distance optics and perfectly works on tissue, potentially paving the way to bulk imaging in turbid samples.
\end{abstract}

It has been demonstrated that an opaque layer may be employed to improve the focusing or imaging capability of an optical system\cite{vellekoop2010exploiting,bertolotti2012non,katz2014non}. Indeed it is possible to fabricate special turbid lenses\cite{van2011scattering} to be placed close to the sample \cite{park2013subwavelength}, and achieving a subdiffraction\cite{yilmaz2015speckle} resolution. This turbid lens-based imaging relies on an high intensity speckle grain\cite{vellekoop2007focusing}, built thought wavefront shaping, and needs for a ``transmission'' geometry, making it not practical for ``in vivo'' measurements of biological systems.

Indeed turbid media generate speckle patterns: macular light structures generated by the interference from scattered or transmitted electromagnetic waves. These are employed, for example, in blind structured illumination microscopy (Blind-SIM), where an image is reconstructed with an improved resolution by sophisticated de-convolution algorithms\cite{mudry2012structured,gustafsson2000surpassing}, typically exploiting the prior knowledge of the sample properties\cite{min2013fluorescent,labouesse2017joint,idier2018superresolution}. Other techniques, such as fluctuation imaging (SOFI)\cite{dertinger2009fast}, rely on higher-order statistical analysis of temporal fluctuations to improve the resolution.  In SOFI  resolution is ultimately limited by the size of the fluctuating item \cite{kim2015superresolution} ( typically the fluorescent molecule or a single speckle grain) even if the practical limit is much higher, due the very high number of frames required to achieve sufficient statistics. In a typical blind-SIM experiment indeed, the speckle grain size is indeed limited by the NA of the illumination and collection optics, which imposes a threshold on the maximum resolution achievable with linear techniques.

On the other hand, in the vicinity of a strongly scattering sample, speckles grains may be much smaller than the illumination point spread function (PSF)\cite{vellekoop2010exploiting}.  These ``\emph{embedded speckles}'', living inside or in the immediate vicinity of a diffusive sample, can reach a size of $\lambda/2\mathfrak{n}$, with $\mathfrak{n}$ being the refractive index of the medium and $\lambda$ the wavelength\cite{goodman2007speckle,carminati2010subwavelength,apostol2003spatial,park2013subwavelength}. As a result, the fluorescent molecules labelling a sample placed in the proximity of a strongly scattering medium, will be excited by small speckles whose size is independent on the collection and illumination optics.


In this work, we describe an illumination strategy working together with an ``opaque mounting medium'' (OMM, see below)  and a custom deconvolution algorithm to improve the imaging resolution by a factor of $\sim$2.5. Our approach exploits the properties of strongly scattering materials placed in the vicinity of the fluorescent samples under analysis. Moreover, a backscattering (reflection) rather than a typical transmission geometry\cite{vellekoop2010exploiting}, is used to enhance the resolution, resulting in an imaging protocol which is particularly advantageous when long working distance optics is needed and transmission geometry is forbidden.

\begin{figure}
\includegraphics[width= 0.7\columnwidth, clip=true] {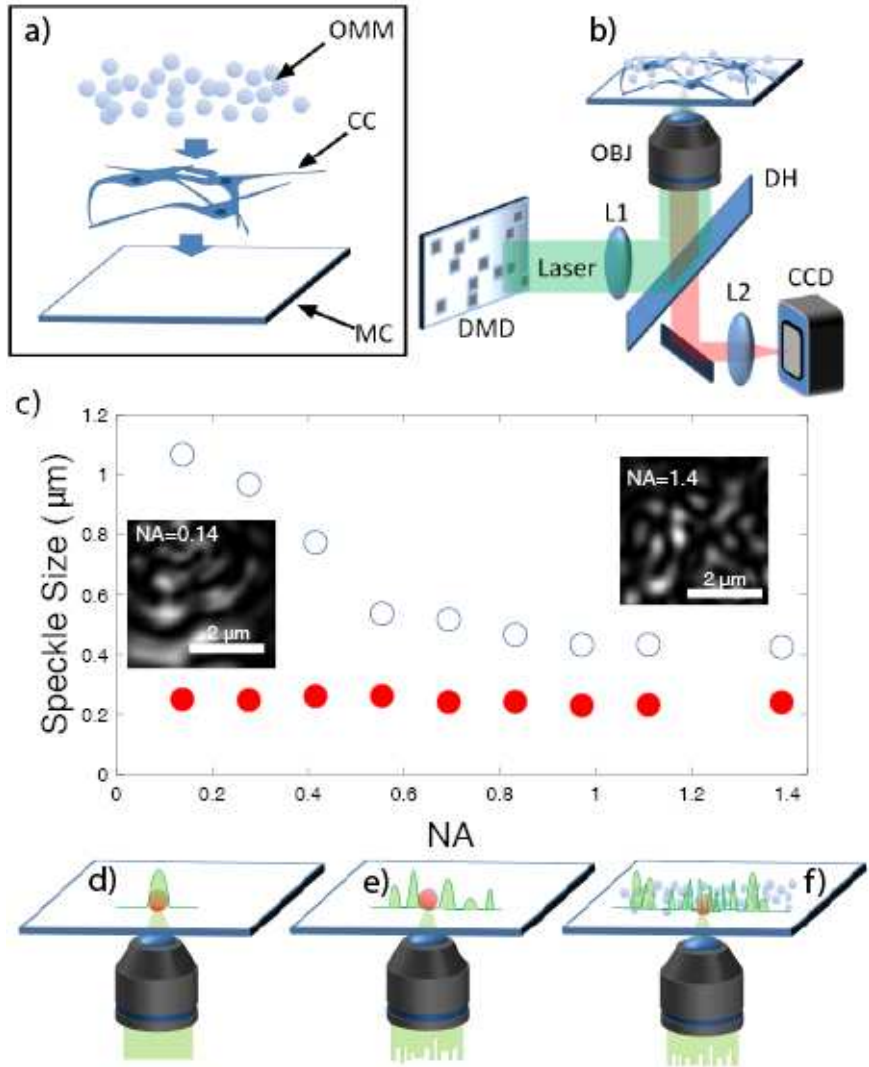}
\caption{\label{SpecklesProps} a) A sketch of the opaque mounting medium (OMM). CC: cell culture. MC:  microscopy coverslip. b) Experimental Setup. The wavefront from a laser source (532 nm) is modulated by a digital micromirror device (DMD) that generates a speckled beam. An image of the DMD plane is demagnified on the sample plane through a telescope composed by lens L1 and the illumination collection objective OBJ. Fluorescent light from the sample is collected through a dichroic mirror (DH) and imaged on a CCD camera through a lens L2. c) Measured speckle size as a function of the NA. Open circles are relative to a flat mirror sample. Full circles are relative to biological sample covered with the OMM. Panels d) e) and f) show three possible illuminations configurations. In a standard illumination (panel d), the focus spot size is defined by the illumination numerical aperture. In panel e), the focal spot size is the same as d) but the illumination is speckled due to the input scrambled wavefront. In f), the speckle size is smaller than the objective PSF due to the presence of the scattering strength of the disordered material given by the OMM. }
\end{figure}

To obtain speckles of size smaller than that defined by the illumination geometry, we exploited the experimental setup shown in Fig. \ref{SpecklesProps}. The OMM backscatters illumination light, generating a contribution with spatial frequencies higher than that of the bare illumination and thus producing smaller speckles. This effect is demonstrated in the plot of Fig. \ref{SpecklesProps}c, where we measured the size of the speckle grains (obtained from the FWHM of the autocorrelation function of the intensity profile) as a function of the effective NA of the illumination beam. To control the illumination NA, we introduced an iris of controllable diameter size $D$ between $L1$ and $DH$ and measured the speckle patterns as a function of  $D$. On the other hand, the collection numerical aperture was kept constant to provide a nominal spatial resolution of $\sim$180 nm.
The experiment has been performed in two configurations. In the first (blue markers in Fig. \ref{SpecklesProps}c ), we measured the size of the speckles generated by the $DMD$ + $OBJ$ system  without disorder,  i.e. using a flat mirror as a sample.  In the second, we measured the speckle size on a biological sample, mounted with our OMM (red markers in Fig. \ref{SpecklesProps}c ). Data show that when the disordered OMM is exploited, the speckles size results to be independent on the illumination NA. This concept, (depicted schematically in panels \ref{SpecklesProps}d-e-f) is at the hearth of the Scattering Assisted Imaging (SAI), in which we can take advantage of the smaller dimension of the illuminating speckles to improve the imaging resolution. In our experimental conditions we retrieved an average speckle size $S$ of 240 $\pm$ 10 nm.

\begin{figure}
\includegraphics[width= \columnwidth, clip=true] {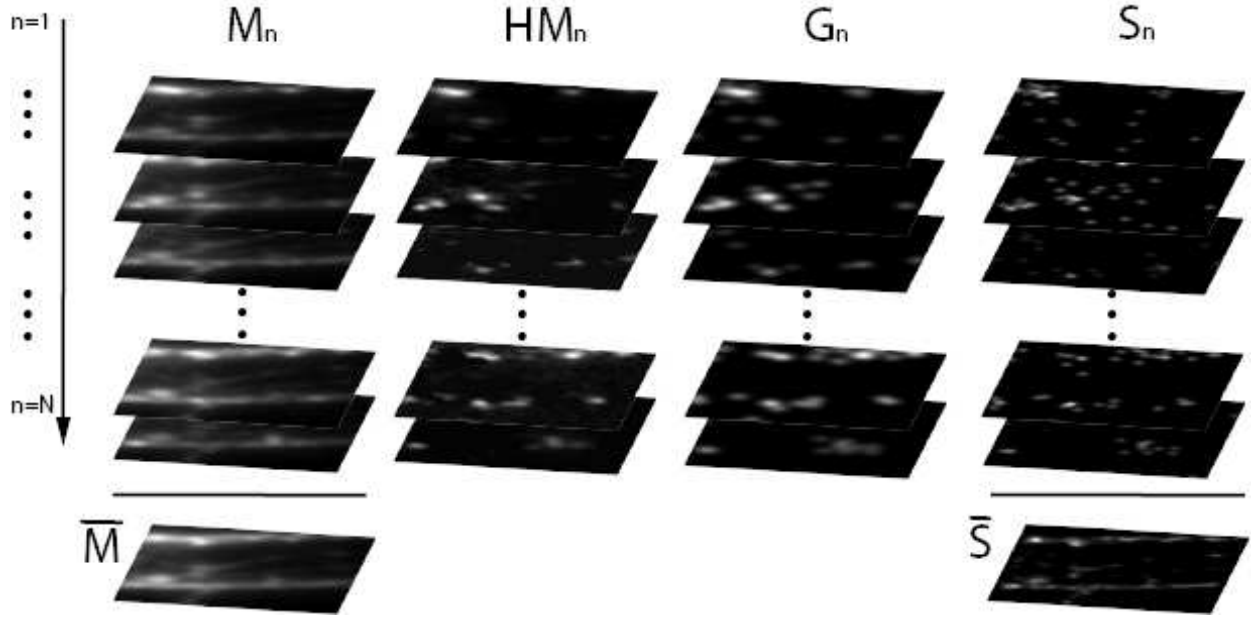}
\caption{\label{Method} $N$ fluorescence images ($M_n$) of a neuron culture stained with Tubulin (see methods) obtained with the illumination patterns $I_n$ are shown in the first column on the left. Summing all contributions, we obtain the average frame $\overline{M}$   (pile bottom). The high intensity part of the fluorescence frame is obtained by subtracting the average frame and considering the positive part of the result. $HM_n$ are reported in the second pile of frames. Applying our gradient descent algorithm, we obtained the $G_n$ (shown in the third column from the left) by minimizing the cost function $F$. The frames shown in right column report the retrieved $S_{n}$. The high resolution image is obtained as the average of the  $ S_{n}$:  $\overline{S}$.}
\end{figure}

Despite that the size of speckle grains at the sample plane is defined by the scattering properties of the OMM rather than the illumination optics, the use of a finite lens in collection would still limit the final imaging resolution of the system due to convolution with the collection PSF. To exceed this limitation, we implemented a custom deconvolution algorithm (Fig. \ref{Method} and see \textbf{see methods}) relying on the prior knowledge of the speckle size.

This process, which is at the heart of SAI, relies on the localization and the intensity assessment of the speckle grains originating the signal through a gradient descent algorithm that minimizes the differences between the target and the guessed distribution of speckle grains. Instead of using the whole information, we restrict our analysis to the high intensity part of the signal which is originated by speckle grains of high intensity (HIGs ). This approach, which relies on a sparser dataset, has the advantage in providing a much faster and more reliable convergence( see \textbf{see methods and supplementary materials}).
This result is directly connected to the fact that in general the $M_n$ fluorescent signal may be generated by a huge number of speckle configurations. On the other hand, $HM_n$ is originated by HIGs with intensity higher than the average value. These HIGs are rare and spatially spars light structures \textbf{ (see supplementary materials)}, thus the deconvolution algorithm requires significantly lower iterations and yields a higher localization accuracy.

\begin{figure}
\includegraphics[width= \columnwidth, clip=true] {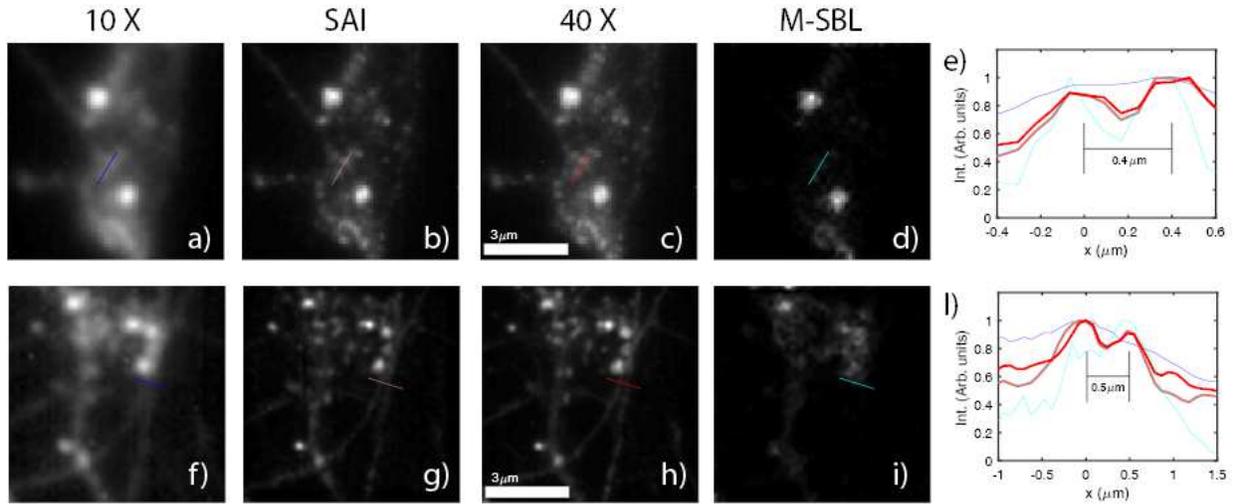}
\caption{\label{Data}  Panels a) and f): low resolution (NA=0.25, nominal resolution $R$=1.1 $\mu m$) fluorescence images obtained averaging N=600 fluorescent frames $M_n$. Panels b) and g): images obtained with SAI with N=600. Panels c) and h): images obtained with high numerical aperture objective (NA=0.75, nominal resolution $R$=360 $\mu m$).Panels d) and i): images obtained with M-SBL reconstruction algorithms with the same N. Panels e) and l): Intensity profiles along the lines highlighted in the previous panels. The color code is the following: Deep blue is for the low fluorescence image; Pink is for  SAI;  Cyan is for  M-SBL; Red is relative to the high numerical aperture objective. }
\end{figure}

Fig. \ref{Data} shows the results obtained applying SAI to a fixed neuron network sample (see \cite{beaudoin2012culturing} and methods). In particular, Figures \ref{Data}a and \ref{Data}f show images from two distinct fields of view obtained averaging $N$=600 speckle frames measured with a 10X (NA=0.25) objective. These images would be equivalent to those obtained in an epi-fluorescence microscope illuminating the sample with a uniform illumination. In turn, Fig. \ref{Data}b and \ref{Data}g show images obtained using SAI on the same field of views. As a reference we further report images obtained with a high numerical aperture (NA=0.75) objective (Fig. \ref{Data}c and \ref{Data}h). Intensity profiles reported in Fig. \ref{Data}e) and \ref{Data}l) shows a resolution of 0.4 $\mu m$ with an objective which has a nominal resolution of 1.1 $\mu m$  thus providing a resolution enhancement of 2.75. As a comparison, we illustrate the same frames analyzed with one of the most promising algorithms for the blind-SIM images deconvolution: the multiple sparse Bayesian learning  (M-SBL)\cite{min2013fluorescent} based on compressive sensing  \cite{donoho2006compressed,katz2009compressive}.  We adapted a recent version of the algorithm \cite{hojman2017photoacousticlink} to our dataset: the results are reported in Fig. \ref{Data}d) and \ref{Data}i). While the algorithm catches a resolution similar to that obtained with SAI, the samples appear altered with respect to the ground truth. This difference may be due to the requirement of a sparse fluorescence arrangement for the M-SBL \cite{hojman2017photoacoustic}. SAI resolution is  also confirmed  by measuring manometric beads\textbf{ (see supplementary materials).}

\begin{figure}
\includegraphics[width= \columnwidth, clip=true]  {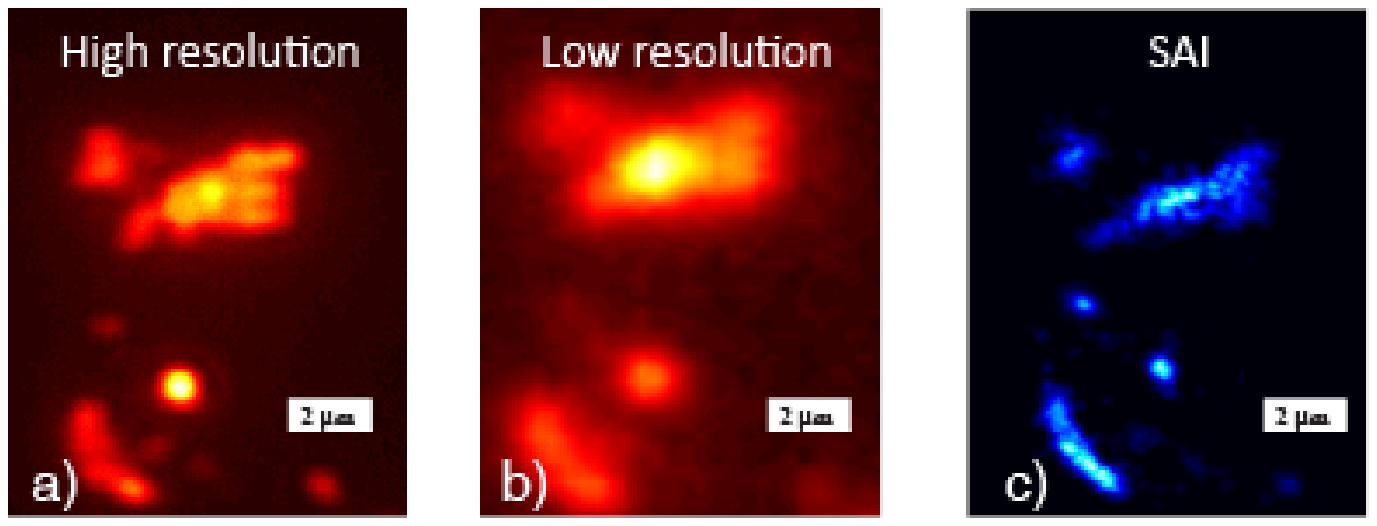}
\caption{ Images of amyloid plaques in a 250 $\mu$m brain slice obtained in a standard configuration with a high NA=0.75 (a) and a low NA=0.25 objective (b). c) Reconstructed SAI image obtained with the low resolution objective lens. SAI achieves  resolution of 0.6 $\pm$ 0.04 $\mu m$ (against a nominal resolution of 1.1 $\mu m$  of the collection objective). We note that despite a measured speckle size $S$ of 350 nm, we retrieve a lower resolution due to the strong sample autofluorescence which is decreasing the effective signal to background ratio in the tissue.  \label{SAS}}
\end{figure}

In the results described above, the scattering was introduced by modifying the optical properties of the mounting medium, thus somehow acting on the sample (even without affecting the fluorescence distribution). To demonstrate the effectiveness of the technique on a naturally scattering  sample, i.e. in a condition in which no further action is needed during the preparation stage, we employed a semi-transparent biological tissue in the form of brain slices containing Amyloid Beta plaques tagged with a fluorescent marker. We used slices derived from a mouse model of the Alzheimer's disease (\textbf{see supplementary materials}), presenting Amyloid Beta plaques. High resolution (obtained with NA=0.75), low resolution (obtained with NA=0.25) and SAI images (obtained with the same NA=0.25 objective) of the Amyloid Beta plaques are reported in Fig. \ref{SAS}l, m and n respectively. Results demonstrate the feasibility of SAI imaging even in biological tissues that naturally introduce a wide range of \textbf{k}-vectors from intrinsic scattering.

In summary, SAI provides an increased resolution without the need of high power laser beams or pulsed sources, while blinking or photo switchable fluorescent chemical compounds are unnecessary. The relative simplicity of the experimental setup makes the proposed approach very general as it can be applied to any fluorescence imaging scheme. Since the resolution limit is driven by the size of the speckle grain, the proposed technique shows its maximum advantage in experiments where high resolution is needed together with a long working distance, a condition that is currently forbidden by the physical NA of the illumination and collection optics. Involving minimal optical power, the scattering assisted imaging  can be a powerful tool for the investigation of systems with low damage threshold, and may be easily exported to \textit{in-vivo} investigation of sensible tissues such as the human retina.

\begin{methods}

\subsection{Experimental setup for speckle measurement and Opaque Mounting Medium}
The proposed configuration is based on a standard imaging scheme in which the illumination is modulated by a DMD generating a scrambled (speckled) wavefront. To exceed the speckle size limit defined by the collection-illumination optics, we exploited the OMM: we embedded a (barely scattering) cell culture in the strongly scattering mounting medium. The opaque mounting medium is a transparent gel embedding the  sample (typically a cell culture or a tissue slice) stained for immunofluorescence experiments \cite{ImpMountingMedium}. Our OMM is realized mixing a standard mounting medium for fluorescence (Agilent-dako fluorescent medium S3023) with Zinc Oxide powder (Zinc Oxide nanopowder $<100nm$ Sigma Aldrich 544906) in a 2 Molar Solution.

\subsection{Reconstruction Algorithm}

In general the fluorescence pattern $M_{n=1....N}$ retrieved with an unknown speckle illumination generated by many successive patterns $I_{n=1....N}$ is given by
\begin{equation}
M_n=(I_n\rho)\bullet h +\epsilon
\label{eq1}
\end{equation}
where $\rho$ is the (fixed) distribution of the fluorophores and $h$ is the PSF of the collection optics while $\bullet$ is the convolution product operator and $\epsilon$ is a noise term. The $n_{th}$ illumination pattern is obtained by randomly orienting the DMD micromirrors, obtaining a total of $N$ fluorescence images. By averaging over all the $N$ speckle frames we obtain the average fluorescence frame $\overline{M}$. By subtracting to each frame the average frame and taking the positive part (indicated by $^+$) of the signal we isolate the part of the signal defined by $HM_n=(M_n-\overline{M})^+$, which is originated by speckles of high intensity. Taking into account eq. \ref{eq1}, we obtain
\begin{equation}
HM_n=((I_n\rho)\bullet h-\overline{M})^+ +\epsilon.
\end{equation}
In fully developed speckles, high intensity grains are (on average) rare and sparse because the intensity probability density function is exponentially decreasing \textbf{(see supplementary materials and \cite{goodman2007speckle})}. Indeed we exploit the $HM_n$ dataset to extract information about the underlying fluorescence distribution. Our hypothesis is that $HM_n$ is generated by a superposition of Gaussian light structures of FWHM $S$  convoluted with the collection optics PSF:
\begin{equation}
HM_n\sim G_n=(S_n)\bullet h +\epsilon
\end{equation}
with
\begin{equation}
S_n=\sum_{k=1:K_n} S_{nk}=\sum_{k=1:K_n} A_{nk} \exp(  (-(\mathbf{r}-\mathbf{R}_{nk})^2)/(2S^2))
\end{equation}
where $\mathbf{r}$ is the coordinate vector in the image plane and $A_{nk}$, $R_{nk}$ and $K_n$ are the intensity, center and total number of the speckle grains respectively that  have to be determined.

To find the best distribution of Gaussians producing the target signal $HM_n$ we implement a proximal gradient descent algorithm \cite{yeh2017structured,parikh2014proximal} minimizing the cost function
\begin{equation}
F=| HM_n-G_n|^2,
\end{equation}
which measures the distance between the target $HM$ and  $G_n$.
Once the pattern $G_n$ is initialized by randomly casting both the intensity $A_k$ and the center $R_k$ of the Gaussians, the algorithm undergoes an optimization process in which a random change of the two parameters is accepted if $F$ is diminished. The procedure is repeated for all the acquired frames and the final high resolution image $\overline{S}$ is obtained by averaging all $S_n$ obtained from the gradient descent procedure as shown in Fig. \ref{Method}.

\subsection{Reliability of the Reconstruction algorithm}
To characterize the reliability of the reconstruction algorithm we measured the degree of similitude $Q$ between the result of two independent gradient descent minimizations $a$ and $b$ ( $Q=\int \mathfrak{N}_{n}^{a}\mathfrak{N}_{n}^{b} dr$, where $\mathfrak{N}^{a}_n= \alpha S^{a}_n$ with $\alpha$ a normalization factor chosen such that  $\int \mathfrak{N}^{a}_n*\mathfrak{N}^{a}_n dr=1$ ). We found that results are very similar  ($Q$ $\sim$ 0.86) if the gradient descent is performed on the high intensity part of the data  ($HM_n$). On the other hand, a degree of similitude of $Q$ $\sim$ 0.35 is found if the original dataset $M_n$ is treated in the same manner.

\subsection{Preparation of the labeled biological sample }
\textbf{Primary cortical neurons} . Primary neuronal cultures were prepared from B6/129 early post-natal (P0-P1) mouse cortex according the protocol previously described in reference 21 of the main paper. Briefly, cortices were isolated from brains and they were dissociated by 20 min incubation in 0.25\% trypsin (15090046, Gibco, Thermo Fisher Scientific) at 37$^\circ$C, 5 min in 0.03\% DNase (000010, Sigma-Aldrich) at RT and mechanically triturated with a fire-polished Pasteur pipette. Cells were plated on poly-l-lysine-coated glass coverslips, and maintained in Neurobasal (21103049, Gibco, Thermo Fisher Scientific) supplemented with 2\% B27 (17504044, Gibco, Thermo Fisher Scientific), 1\% L-Glutamine 200mM (59202C, Sigma-Aldrich) and 1\% Penicillin-Streptomycin (P4333, Sigma-Aldrich). Cells were cultured in controlled environment, with a humidified atmosphere containing 5\% CO$_2$ at 37 $^\circ$CC. Half of the grow medium was changed every 2 days.\\

\textbf{Immunofluorescence assay.} After 14 days, neuronal primary cultures were stained for the detection of BetaIII-tubulin. Briefly, dishes were fixed in 4\% PFA 15' and, after 3' permeabilization in 0,1\% Triton X-100 and 1h blocking in 1\% BSA, they were incubated with primary antibody (T2200, Sigma Aldrich, 1:1000 in 0,1\% BSA). After 18h and 3 washes in PBS, secondary antibody (anti-rabbit Alexa Fluor 532, \#A-11009, Thermo Fisher Scientific, 1:500 in 0,1\% BSA) was added for 45' and coverslips were then dehydrated by consecutive 2 minute washes in increasing doses of ethanol (30-50-70-90\%) and then mounted with our opaque mounting medium.\\

\textbf{Opaque mounting medium.}  Our OMM is realized mixing a standard mounting medium for fluorescence (Agilent-dako fluorescent medium S3023) with Zinc Oxide powder (Zinc Oxide nanopowder $<100nm$ Sigma Aldrich 544906) in a 2 Molar Solution. The OMM backscatters illumination light, generating a contribution with spatial frequencies higher than that of the bare illumination and thus producing smaller speckles. We depose by drop-cast 20 $\mu$l of OMM on the microscopy coverslip which is hosting the culture, which is then squeezed with a second microscopy coverglass which is then sealed, thus sandwiching the culture between the coverglass and the OMM.\\

\textbf{Brain Slice preparation} All experiments on animals were conducted in conformity with European Directive 2010/63/EU and the Italian D.lg. 4.05.2014 and all methods were carried out in accordance with relevant guidelines and regulations. One-year old 3xTg-AD mice were euthanized and transcardially perfused with cold Phosphate Buffered Saline (PBS) solution. 300µm thick slices were obtained with a vibratome. Slices were fixed in a 4\% Paraformaldeide solution for 16 hours at 4$^\circ$C and then processed for the free-floating immunostaining. Slices were treated with a solution of 70\% formic acid for 30 minutes to reveal antigen and then blocked with 3\% goat serum and 0,3\% Triton X-100 in PBS for 1 hour; Amyloid Beta-recognizing primary antibody (803001, Biolegend) was added 1:100 in a solution of 1\% goat serum and 0,1\% Triton X-100 in PBS at 4C for 16 hours in continuous agitation. After 3 washes in PBS, Goat anti-Mouse IgG (H\&L) Coated Fluorescent Nile Red secondary antibody (MFP-0556-5, Spherotech) was added for 1h and then the last 3 washes in PBS were performed. Stained brain slices were mounted on a slide with a fluorescent mounting medium (Agilent-dako fluorescent medium S3023) and covered with a coverslip.\\

\end{methods}


\begin{thebibliography}{10}
\expandafter\ifx\csname url\endcsname\relax
  \def\url#1{\texttt{#1}}\fi
\expandafter\ifx\csname urlprefix\endcsname\relax\def\urlprefix{URL }\fi
\providecommand{\bibinfo}[2]{#2}
\providecommand{\eprint}[2][]{\url{#2}}

\bibitem{abbe1873beitrage}
\bibinfo{author}{Abbe, E.}
\newblock \bibinfo{title}{Beitr{\"a}ge zur theorie des mikroskops und der
  mikroskopischen wahrnehmung}.
\newblock \emph{\bibinfo{journal}{Archiv f{\"u}r mikroskopische Anatomie}}
  \textbf{\bibinfo{volume}{9}}, \bibinfo{pages}{413--418}
  (\bibinfo{year}{1873}).

\bibitem{rayleigh1903theory}
\bibinfo{author}{Rayleigh, L.}
\newblock \bibinfo{title}{On the theory of optical images, with special
  reference to the microscope}.
\newblock \emph{\bibinfo{journal}{Journal of the Royal Microscopical Society}}
  \textbf{\bibinfo{volume}{23}}, \bibinfo{pages}{474--482}
  (\bibinfo{year}{1903}).

\bibitem{sheng2006introduction}
\bibinfo{author}{Sheng, P.}
\newblock \emph{\bibinfo{title}{Introduction to wave scattering, localization
  and mesoscopic phenomena}}, vol.~\bibinfo{volume}{88}
  (\bibinfo{publisher}{Springer Science \& Business Media},
  \bibinfo{year}{2006}).

\bibitem{bertolotti2012non}
\bibinfo{author}{Bertolotti, J.} \emph{et~al.}
\newblock \bibinfo{title}{Non-invasive imaging through opaque scattering
  layers}.
\newblock \emph{\bibinfo{journal}{Nature}} \textbf{\bibinfo{volume}{491}},
  \bibinfo{pages}{232} (\bibinfo{year}{2012}).

\bibitem{vellekoop2010exploiting}
\bibinfo{author}{Vellekoop, I.~M.}, \bibinfo{author}{Lagendijk, A.} \&
  \bibinfo{author}{Mosk, A.}
\newblock \bibinfo{title}{Exploiting disorder for perfect focusing}.
\newblock \emph{\bibinfo{journal}{Nature photonics}}
  \textbf{\bibinfo{volume}{4}}, \bibinfo{pages}{320--322}
  (\bibinfo{year}{2010}).

\bibitem{park2013subwavelength}
\bibinfo{author}{Park, J.-H.} \emph{et~al.}
\newblock \bibinfo{title}{Subwavelength light focusing using random
  nanoparticles}.
\newblock \emph{\bibinfo{journal}{Nature photonics}}
  \textbf{\bibinfo{volume}{7}}, \bibinfo{pages}{454--458}
  (\bibinfo{year}{2013}).

\bibitem{katz2014non}
\bibinfo{author}{Katz, O.}, \bibinfo{author}{Heidmann, P.},
  \bibinfo{author}{Fink, M.} \& \bibinfo{author}{Gigan, S.}
\newblock \bibinfo{title}{Non-invasive single-shot imaging through scattering
  layers and around corners via speckle correlations}.
\newblock \emph{\bibinfo{journal}{Nature photonics}}
  \textbf{\bibinfo{volume}{8}}, \bibinfo{pages}{784} (\bibinfo{year}{2014}).

\bibitem{van2011scattering}
\bibinfo{author}{Van~Putten, E.} \emph{et~al.}
\newblock \bibinfo{title}{Scattering lens resolves sub-100 nm structures with
  visible light}.
\newblock \emph{\bibinfo{journal}{Physical review letters}}
  \textbf{\bibinfo{volume}{106}}, \bibinfo{pages}{193905}
  (\bibinfo{year}{2011}).

\bibitem{yilmaz2015speckle}
\bibinfo{author}{Yilmaz, H.} \emph{et~al.}
\newblock \bibinfo{title}{Speckle correlation resolution enhancement of
  wide-field fluorescence imaging}.
\newblock \emph{\bibinfo{journal}{Optica}} \textbf{\bibinfo{volume}{2}},
  \bibinfo{pages}{424--429} (\bibinfo{year}{2015}).

\bibitem{vellekoop2007focusing}
\bibinfo{author}{Vellekoop, I.~M.} \& \bibinfo{author}{Mosk, A.}
\newblock \bibinfo{title}{Focusing coherent light through opaque strongly
  scattering media}.
\newblock \emph{\bibinfo{journal}{Optics letters}}
  \textbf{\bibinfo{volume}{32}}, \bibinfo{pages}{2309--2311}
  (\bibinfo{year}{2007}).

\bibitem{mudry2012structured}
\bibinfo{author}{Mudry, E.} \emph{et~al.}
\newblock \bibinfo{title}{Structured illumination microscopy using unknown
  speckle patterns}.
\newblock \emph{\bibinfo{journal}{Nature Photonics}}
  \textbf{\bibinfo{volume}{6}}, \bibinfo{pages}{312--315}
  (\bibinfo{year}{2012}).

\bibitem{gustafsson2000surpassing}
\bibinfo{author}{Gustafsson, M.~G.}
\newblock \bibinfo{title}{Surpassing the lateral resolution limit by a factor
  of two using structured illumination microscopy}.
\newblock \emph{\bibinfo{journal}{Journal of microscopy}}
  \textbf{\bibinfo{volume}{198}}, \bibinfo{pages}{82--87}
  (\bibinfo{year}{2000}).

\bibitem{min2013fluorescent}
\bibinfo{author}{Min, J.} \emph{et~al.}
\newblock \bibinfo{title}{Fluorescent microscopy beyond diffraction limits
  using speckle illumination and joint support recovery}.
\newblock \emph{\bibinfo{journal}{Scientific reports}}
  \textbf{\bibinfo{volume}{3}}, \bibinfo{pages}{2075} (\bibinfo{year}{2013}).

\bibitem{labouesse2017joint}
\bibinfo{author}{Labouesse, S.} \emph{et~al.}
\newblock \bibinfo{title}{Joint reconstruction strategy for structured
  illumination microscopy with unknown illuminations}.
\newblock \emph{\bibinfo{journal}{IEEE Transactions on Image Processing}}
  \textbf{\bibinfo{volume}{26}}, \bibinfo{pages}{2480--2493}
  (\bibinfo{year}{2017}).

\bibitem{idier2018superresolution}
\bibinfo{author}{Idier, J.} \emph{et~al.}
\newblock \bibinfo{title}{On the superresolution capacity of imagers using
  unknown speckle illuminations}.
\newblock \emph{\bibinfo{journal}{IEEE Transactions on Computational Imaging}}
  \textbf{\bibinfo{volume}{4}}, \bibinfo{pages}{87--98} (\bibinfo{year}{2018}).

\bibitem{dertinger2009fast}
\bibinfo{author}{Dertinger, T.}, \bibinfo{author}{Colyer, R.},
  \bibinfo{author}{Iyer, G.}, \bibinfo{author}{Weiss, S.} \&
  \bibinfo{author}{Enderlein, J.}
\newblock \bibinfo{title}{Fast, background-free, 3d super-resolution optical
  fluctuation imaging (sofi)}.
\newblock \emph{\bibinfo{journal}{Proceedings of the National Academy of
  Sciences}} \textbf{\bibinfo{volume}{106}}, \bibinfo{pages}{22287--22292}
  (\bibinfo{year}{2009}).

\bibitem{kim2015superresolution}
\bibinfo{author}{Kim, M.}, \bibinfo{author}{Park, C.},
  \bibinfo{author}{Rodriguez, C.}, \bibinfo{author}{Park, Y.} \&
  \bibinfo{author}{Cho, Y.-H.}
\newblock \bibinfo{title}{Superresolution imaging with optical fluctuation
  using speckle patterns illumination}.
\newblock \emph{\bibinfo{journal}{Scientific reports}}
  \textbf{\bibinfo{volume}{5}}, \bibinfo{pages}{16525} (\bibinfo{year}{2015}).

\bibitem{goodman2007speckle}
\bibinfo{author}{Goodman, J.~W.}
\newblock \emph{\bibinfo{title}{Speckle phenomena in optics: theory and
  applications}} (\bibinfo{publisher}{Roberts and Company Publishers},
  \bibinfo{year}{2007}).

\bibitem{carminati2010subwavelength}
\bibinfo{author}{Carminati, R.}
\newblock \bibinfo{title}{Subwavelength spatial correlations in near-field
  speckle patterns}.
\newblock \emph{\bibinfo{journal}{Physical Review A}}
  \textbf{\bibinfo{volume}{81}}, \bibinfo{pages}{053804}
  (\bibinfo{year}{2010}).

\bibitem{apostol2003spatial}
\bibinfo{author}{Apostol, A.} \& \bibinfo{author}{Dogariu, A.}
\newblock \bibinfo{title}{Spatial correlations in the near field of random
  media}.
\newblock \emph{\bibinfo{journal}{Physical review letters}}
  \textbf{\bibinfo{volume}{91}}, \bibinfo{pages}{093901}
  (\bibinfo{year}{2003}).

\bibitem{beaudoin2012culturing}
\bibinfo{author}{Beaudoin~III, G.~M.} \emph{et~al.}
\newblock \bibinfo{title}{Culturing pyramidal neurons from the early postnatal
  mouse hippocampus and cortex}.
\newblock \emph{\bibinfo{journal}{Nature protocols}}
  \textbf{\bibinfo{volume}{7}}, \bibinfo{pages}{1741} (\bibinfo{year}{2012}).

\bibitem{donoho2006compressed}
\bibinfo{author}{Donoho, D.~L.}
\newblock \bibinfo{title}{Compressed sensing}.
\newblock \emph{\bibinfo{journal}{IEEE Transactions on information theory}}
  \textbf{\bibinfo{volume}{52}}, \bibinfo{pages}{1289--1306}
  (\bibinfo{year}{2006}).

\bibitem{katz2009compressive}
\bibinfo{author}{Katz, O.}, \bibinfo{author}{Bromberg, Y.} \&
  \bibinfo{author}{Silberberg, Y.}
\newblock \bibinfo{title}{Compressive ghost imaging}.
\newblock \emph{\bibinfo{journal}{Applied Physics Letters}}
  \textbf{\bibinfo{volume}{95}}, \bibinfo{pages}{131110}
  (\bibinfo{year}{2009}).

\bibitem{hojman2017photoacousticlink}
\bibinfo{author}{Hojman, E.}
\newblock \emph{\bibinfo{title}{Photoacoustic object recovery through M-SBL
  [Data set]}} (\bibinfo{publisher}{Zenodo,
  https://doi.org/10.5281/zenodo.268701}, \bibinfo{year}{2017}).

\bibitem{hojman2017photoacoustic}
\bibinfo{author}{Hojman, E.} \emph{et~al.}
\newblock \bibinfo{title}{Photoacoustic imaging beyond the acoustic
  diffraction-limit with dynamic speckle illumination and sparse joint support
  recovery}.
\newblock \emph{\bibinfo{journal}{Optics express}}
  \textbf{\bibinfo{volume}{25}}, \bibinfo{pages}{4875--4886}
  (\bibinfo{year}{2017}).

\bibitem{ImpMountingMedium}
\bibinfo{author}{Lennette, D.~A.}
\newblock \bibinfo{title}{An improved mounting medium for immunofluorescence
  microscopy}.
\newblock \emph{\bibinfo{journal}{American Journal of Clinical Pathology}}
  \textbf{\bibinfo{volume}{69}}, \bibinfo{pages}{647--648}
  (\bibinfo{year}{1978}).

\bibitem{yeh2017structured}
\bibinfo{author}{Yeh, L.-H.}, \bibinfo{author}{Tian, L.} \&
  \bibinfo{author}{Waller, L.}
\newblock \bibinfo{title}{Structured illumination microscopy with unknown
  patterns and a statistical prior}.
\newblock \emph{\bibinfo{journal}{Biomedical optics express}}
  \textbf{\bibinfo{volume}{8}}, \bibinfo{pages}{695--711}
  (\bibinfo{year}{2017}).

\bibitem{parikh2014proximal}
\bibinfo{author}{Parikh, N.}, \bibinfo{author}{Boyd, S.} \emph{et~al.}
\newblock \bibinfo{title}{Proximal algorithms}.
\newblock \emph{\bibinfo{journal}{Foundations and Trends{\textregistered} in
  Optimization}} \textbf{\bibinfo{volume}{1}}, \bibinfo{pages}{127--239}
  (\bibinfo{year}{2014}).

\end{thebibliography}


\begin{addendum}
 \item \emph{Aknowledgements}:M. L. acknowledges ``Fondazione CON IL SUD,'' Grant ``Brains2south'', Project ``Localitis''..
 \item[Competing Interests] The authors declare that they have no
competing financial interests.
 \item[Correspondence] Correspondence and requests for materials should be addressed to ML(email: : marco.leonetti@iit.it).
\end{addendum}


\end{document}